\documentclass[9pt,twocolumn,twoside]{article}

\usepackage[utf8]{inputenc}
\usepackage[letterpaper,left=0.8in,top=0.5in,right=0.8in,bottom=1.0in]{geometry}
\usepackage{hyperref,amsbsy,amsmath,amsthm,amsfonts,mathrsfs,float}
\usepackage{graphicx,subfigure,tabularx,bm,soul}
\usepackage{enumitem,amssymb,lipsum,abstract,parskip}
\usepackage[labelfont=bf]{caption}
\usepackage[dvipsnames]{xcolor}
\usepackage{sectsty}
\usepackage[switch]{lineno}
\usepackage{authblk}
\usepackage{physics}
\newlist{todolist}{itemize}{2}
\setlist[todolist]{label=$\square$}
\usepackage{pifont}

\usepackage{bbm,mathtools,cases,multirow}
\usepackage[normalem]{ulem}

\allsectionsfont{\nohang\centering\normalsize\uppercase}
\subsectionfont{\nohang\centering\small}

\newcommand{\lb}{\left(}
\newcommand{\rb}{\right)}

\graphicspath{{Figures/}}

\usepackage{ulem}
\usepackage{lipsum}

\begin{document}

\title{Competing forces of polarization and adhesion generate directional migration bias in a minimal model}

\author[1]{Egun Im}
\author[2]{Ghina Badih}
\author[2]{Laetitia Kurzawa}
\author[3,*]{Andreas Buttensch\"{o}n}
\author[1,4,*]{Calina Copos}

\affil[1]{\small{Department of Biology, Northeastern University}}
\affil[2]{\small{Universit\'e Grenoble-Alpes, CEA, CNRS, INRA, Interdisciplinary Research Institute of Grenoble, CytoMorpho Lab, Grenoble, France}}
\affil[3]{\small{Department of Mathematics and Statistics, University of Massachusetts Amherst}}
\affil[4]{\small{Department of Mathematics, Northeastern University}}
\affil[*]{\small{Co-corresponding authors: andreas.buttenschoen@umass.edu, c.copos@northeastern.edu}}

\date{}

\twocolumn[
  \maketitle
\begin{abstract}
\vspace{2mm}
    Left-right axis specification establishes embryonic laterality through asymmetric signaling cascades originating at the cellular scale. We previously reported the presence of a directionality bias in confined pairs of endothelial (and fibroblast) cells exhibiting persistent circular motion, with cytoskeletal contractility modulating the direction. The relative simplicity of the experimental setup makes it a perfect testing ground for the physical forces that could endow this system with a tunable directional migration bias. We model self-propelling biological cells migrating in response to confinement, polarity, and pairwise repulsive forces. Our framework reproduces three key experimental observations: spontaneous coherent circular movement of confined cell pairs, emergence of directional bias when cells have asymmetric properties, and contractility-modulated switching of the rotation direction. Two key assumptions are required: an internal torque arising from cytoskeletal organization (previously observed in other cellular systems), and an asymmetric polarity response between cells, which introduces a difference in how quickly each cell reorients its migration direction. New experiments on daughter cell pairs support this asymmetry requirement in cellular properties. Tuning the polarity response timescale (or strength) relative to centering forces from confinement and cell-cell adhesion can amplify or reverse the directional migration bias.
\vspace{5mm}
\end{abstract}
]

\fbox{\parbox{0.48\textwidth}{\noindent\textbf{Statement of Significance:} Left-right asymmetry is vital in embryonic development, originating from cells regulated by mechanical and biochemical signaling. Chirality -- directional "handedness"—appears throughout nature, from snail shell coiling to plant root spiraling. Recent experiments showed directionality bias in cellular movement can be established and reversed by varying contractility of confined cell pairs. Using a modeling framework, we demonstrate how physical forces distinguish left from right, endowing systems with tunable directionality bias in confined migration. Individual cells must possess intrinsic bias in their front-rear polarity axis. By varying contractility, cells tune their adherence to their environment or each other. A tug-of-war emerges between orientation biased polarity and centering forces from adhesions. This study posits a minimal framework for biased directional movement.}}


\noindent\textbf{Running title}: Directional bias in cell migration

\section{Introduction}

Chirality, the property that an object cannot be superimposed on its mirror image, is a conserved feature of living organisms with critical implications during embryonic development~\cite{Asai2025,Zhang2024,Davison2016,Coutelis2014,levin2005left,Wood1997}. In the mouse embryo, rotating cilia generate a right-to-left fluid flow fundamental to the organism's left-right asymmetry of organ positioning and shape during development~\cite{Nonaka1998}. Asymmetric organ morphogenesis is believed to be controlled by the differential expression of signaling cascades reviewed in~\cite{levin2005left} --- for example, in both chick and mouse normal embryos, Nodal expression is observed on the left side. There is increasing evidence that these signaling networks, in turn, must follow asymmetric events by their constituent cells~\cite{tee2023actin,ray2018intrinsic,inaki2016cell}. Beyond biological cells, there is a growing interest in chiral active matter, including a variety of biological circle swimmers, such as~\textit{E.coli}~\cite{Lauga2016,DiLeonardo2010,DiLuzio2005,Berg1993}, sperm~\cite{Friedrich2007,Riedel2005}, and magnetotactic bacteria~\cite{Cebers2011,Erglis2007}, as well as synthetic self-propelled chiral particles~\cite{WingChan2024,tenHagen2014}.

\begin{figure}[h!]
    \centering
    \includegraphics[width=0.5\textwidth]{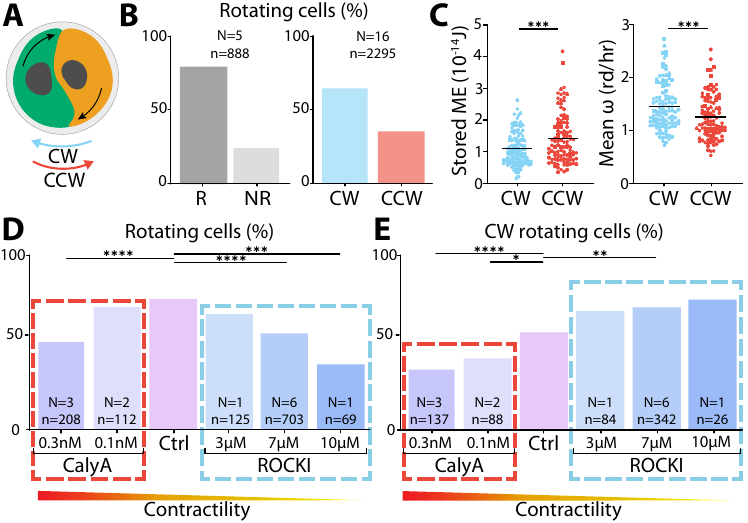}
    \caption{\textbf{Summary of our experimental results in~\cite{Badih2025} on the movement of HUVEC pairs confined to a disk-shaped geometry.} (A) Schematic of the experimental setup illustrating two HUVEC cells confined to a disk-shaped adherent geometry with $R=60~\mu m$. (B) Summary of experimental results illustrating a three-state migratory system: doublets rotating coherently (R) in either the clockwise (CW) or counterclockwise (CCW) direction or non-rotating (NR). (C) The doublets are characterized by an asymmetry in the stored mechanical energy as reported by traction force measurements and in the angular speed; Statistical significance was assessed using an unpaired t-test ($p=0.0087$ for left plot; $p=0.0110$ for right plot). $N=3$ independent experiments, and $n=98$ doublets for both plots. (D) Percentage of rotational doublets and (E) percentage of CW-rotational doublets, both showing directional biases modulated by contractility. Quantified in control versus treated, from decreasing concentrations of CalyA to increasing concentrations of ROCKI. Statistical significance was assessed using Chi-squared test (Fischer's exact; Significance testing: *$=0.01238$; **$=0.0087$; ***$=0.0007$; ****$\leq0.0001$). $N$ indicates the number of individual experiments and $n$ the total number of doublets used for quantification.}
    \label{fig:exp_results}
\end{figure}

One of the main unresolved issues arising in both active and biological matter is the ability to identify the macroscopic forces that distinguish left from right in order to form and abolish biased movement. Here, we focus on a minimal experimental setup from our recent work that demonstrates robust control of biased directional movement of a pair of cells in a confined geometry~\cite{Badih2025}. A few important observations were made in this minimal system composed of a pair of human umbilical vein endothelial cells (HUVEC) confined to a disk-shaped fibronectin-coated micropattern (Fig.~\ref{fig:exp_results}A). (1) The doublets spontaneously and persistently rotate, albeit not always (Fig.~\ref{fig:exp_results}B). (2) A mild bias towards the clockwise (CW) direction was observed (Fig.~\ref{fig:exp_results}B). On average, around 80\%-90\% of the HUVEC doublets rotated persistently, with 60\% of those in the CW-direction. (3) CW doublets rotated faster and exerted less force compared to their counterclockwise (CCW) counterparts  (Fig.~\ref{fig:exp_results}C). (4) By using contractility modulating drugs, such as Rho kinase inhibitor (ROCKI) and Calyculin A (CalyA), the bias could be amplified or even reversed (Fig.~\ref{fig:exp_results}D-E). The authors speculate that the speed and direction of rotation are determined by the more contractile cells within the doublets. Even so, a mechanistic understanding of how mechanical forces are integrated to give rise to chiral bias remains unknown. Specifically, no macroscopic force distinguishes left from right, leaving open the question of how directional bias emerges.  

Previous models of the rotations of cell groups have been proposed, some of which explicitly capture biases~\cite{tee2023actin,Rahman2023,wangxu2022,xu2022,erzberger2020mechanochemical} while others do not~\cite{riveline2023,vahabliviczek2023}. In particular, the integrity of the actin network and actin-related proteins has been implicated in the establishment of the chirality of movement, including formins~\cite{tee2015cellular}, alpha-actinin~\cite{TeeBershasky2023, KwongChen2023,ChinWan2018}, myosin~\cite{KwongChen2023}, and associated signaling pathways~\cite{erzberger2020mechanochemical}. None of these theoretical works focus on the modulation of chirality. Motivated by our experimental findings in Badih et al.~\cite{Badih2025}, we address this critical question here.

We begin with a focus on recent experiments performed in~\cite{Badih2025}. Using analytical and computational tools applied to a mathematical model, we recapitulate these experiments to examine the rotational movement patterns of a pair of cells confined to a disk-shaped geometry. Using a friction-dominated minimal system, we can recapitulate unbiased circular movement in this confined geometry, as well as the emergence of CW or CCW biased movement depending on how quickly cells reorient their polarity relative to the mechanical forces from confinement and cell-cell adhesion. Modulating the polarity response, which in our model is related to cellular contractility, can amplify or even reverse the chirality bias, just as observed experimentally. We demonstrate that our model is applicable to other cell types and validate the model findings with new experiments on daughter cell pairs that lack asymmetry in stored mechanical energy. Our study, therefore, provides a minimal description of the macroscopic cellular forces that produce bias.

\section{Mathematical Model}

\begin{figure*}[h!]
    \centering
    \includegraphics[width=\textwidth]{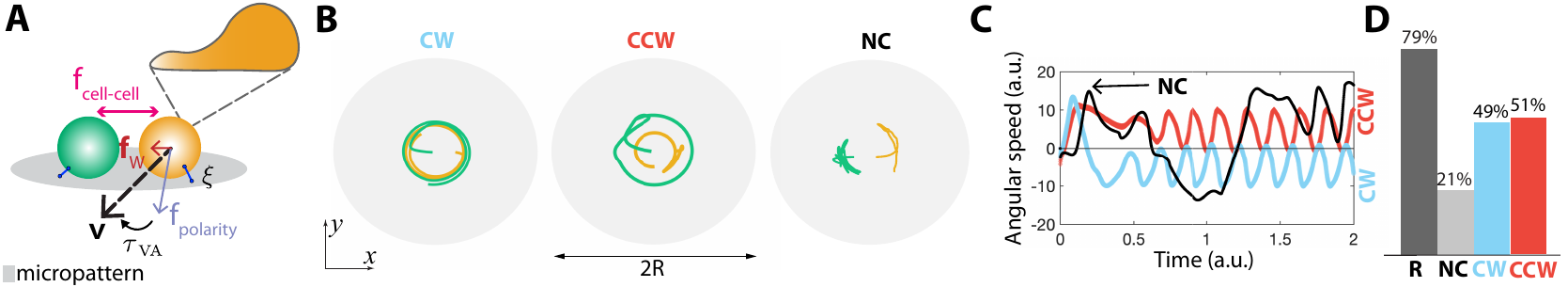}
    \caption{\textbf{Model of a pair of cells confined to a disk and their circular movement dynamics.} (A) Doublet model schematic. (B) Sample CW, CCW, and NC trajectories over one rotational cycle. (C) Angular velocity over~2 arbitrary time units for cells that move persistently in the CW (blue) and CCW (red) directions or move non-coherently by switching directionality (black dots). (D) Averaged number of rotating doublets and their distribution into the three motility states: CW (49\%), CCW (51\%), and NC (21\%) out of 3200 model simulations.}
    \label{fig:doublet_schematic}
\end{figure*}

We build a minimal mathematical model to capture the interplay between cell polarity, contractility, cell-cell adhesion, and spatial confinement (Fig.~\ref{fig:doublet_schematic}).
Each cell is represented by the position of its center-of-mass $\mathbf{x}(t)$; detailed morphology such as the cell membrane, nucleus, and actomyosin cytoskeleton is not explicitly modeled. In the overdamped limit, the forces on each cell balance at every instant, so that a cell moves with a velocity $\mathbf{v}(t)$ determined by:
\begin{equation}
\begin{split}
    \sum \mathbf{f}(\mathbf{x},t) &= 0 \\ \Rightarrow \mathbf{v}^{(i)} = \dot{\mathbf{x}}^{(i)} &= 
    \frac{1}{\xi^{(i)}} \left( \mathbf{f}_\text{polarity}^{(i)} + \gamma_\text{cc}\mathbf{f}_\text{cell-cell}^{(i)} + \gamma_\text{wall}\mathbf{f}_\text{W}^{(i)}\right),
\label{eq:singlet_forcebalance}
\end{split}
\end{equation}
where the superscript identifies which cell is being described.~$\xi$ is the drag coefficient representing friction between the cell and the substrate, which at high adhesion bond turnover rates is effectively equivalent to elastic cell-matrix interactions~\cite{Howard2018}. The forces per unit length acting on the cell are: (1) frictional drag force between the cell and the surface underneath ($\mathbf{f}_\text{drag} = \xi \dot{\mathbf{x}}$), (2) active polarity force arising from front-rear signaling of actin-based protrusions and myosin-based contraction ($\mathbf{f}_\text{polarity}$), (3) cell-cell interaction with coupling strength $\gamma_\text{cc}$ ($\mathbf{f}_\text{cell-cell}$), and (4) a confinement force $\mathbf{f}_\text{W} = -(\cos\theta_W, \sin\theta_W)^T$ directed toward the center of the micropattern, where $\theta_W$ is the angle from the cell's position to the domain center (see Spatial Confinement below), with strength $\gamma_\text{wall}$, which is nonzero only when the polarity strength exceeds the baseline value (Table~S1).

\textbf{Cell polarity.} Migrating cells have an underlying chemical polarization, indicating the areas of the cell that are likely to protrude (``front-like'') and those likely to contract (``rear-like'')~\cite{LeahReview}. This can include an asymmetric distribution of Rho GTPases, with Rac1 activity driving the cell front through lamellipodial extensions and RhoA promoting myosin contractility in the rear. Rather than explicitly modeling the dynamics of one or more Rho GTPases~\cite{Mori2008,Buttenschon2020}, we summarize cell polarity with a single spatiotemporal motility force $\mathbf{f}_\text{polarity}$.
This force
is a vector with direction $\phi$, initially chosen randomly, and magnitude
$\gamma_{\mathrm{pol}}$, a model parameter:
\begin{equation}
    \mathbf{f}_\text{polarity} = \gamma_{\mathrm{pol}} \left(\cos{\phi},\sin{\phi}\right)^{T}.
\label{eq:polarity_a}
\end{equation}
\noindent Over time, the polarity direction $\phi$ gradually rotates toward the cell's current direction
of motion, a mechanism known as velocity alignment~\cite{Camley2014}:
\begin{equation}
    \dot{\phi} = \frac{1}{\tau_\text{VA}}\arcsin{\left[\cos{\phi}\sin{\theta_V} - \sin{\phi}\cos{\theta_V}\right]}. 
\label{eq:polarity_b}
\end{equation}
\noindent Here, $\theta_V$ is the direction of the velocity vector $\mathbf{v}(t)$, computed as $\arctan{(v_y/v_x)}$, and $\tau_\text{VA}$ sets the orientational persistence timescale. When $\tau_{\text{VA}}$ is large, the cell is slow to adjust its polarity and moves with high persistence in its current polarity direction. When $\tau_{\text{VA}}$ is small, the cell rapidly reorients its polarity to match its current velocity. This form is the one originally suggested by~\cite{Szabo2006} and later adapted by~\cite{Camley2014} and others.

\textbf{Spatial confinement.}
The experiments of~\cite{Badih2025} use an adhesive fibronectin-coated micropattern of radius $R$ to effectively confine the cells --- cells can only adhere to the substrate on the micropattern. We incorporate confinement through a reorientation of cell polarity in
Eq.~\eqref{eq:polarity_b}:
\begin{multline}
    \dot{\phi} = \frac{1}{\tau_\text{VA}}\arcsin{\left[\cos{\phi}\sin{\theta_V} - \sin{\phi}\cos{\theta_V}\right]} \\\\ + \frac{1}{\tau_\text{W}}\arcsin{\left[\cos{\phi}\sin{\theta_W} - \sin{\phi}\cos{\theta_W}\right]},
\label{eq:polarity_c}
\end{multline}
where $\theta_W$ is the angle pointing from the cell's position toward the center of the domain. This formulation captures how cells continuously sense the boundary of the micropattern through their cytoskeleton and steer back toward the interior. Unlike an explicit confining force that activates at a fixed distance from the boundary, the reorientation of polarity provides a smooth, biologically motivated response to geometric constraints. In the absence of such a confinement effect, the cells engage in persistent directional motion. We find that this implicit approach successfully recapitulates the experimental findings, though others are possible.

\textbf{Cell-cell interaction.} There is no single established approach for modeling the complex interactions between neighboring cells~\cite{ShiReview2025}. Here, cells interact through volume exclusion: when two cells are closer than a threshold distance, they push each other apart. This repulsive force indirectly influences each cell's polarity through the velocity alignment mechanism described above. In this way, the volume exclusion captures the macroscopic effect of contact inhibition of locomotion (CIL), whereby cells repolarize away from each other upon contact, without explicitly modeling the underlying Rac/Rho signaling dynamics. In Supporting Material, we show that other intercellular couplings, including passive (elastic) interactions, do not impact the qualitative behavior of our findings. The volume exclusion force is described by
\begin{equation}
    \mathbf{f}_\text{cell-cell}^{(i)}
        = \lb \norm{\mathbf{x}_{ij}} - \ell_0 \rb^{-}
        \frac{\mathbf{x}_{ij}}{\norm{\mathbf{x}_{ij}}}
\label{eq:cell-cellforce}
\end{equation}
where $\mathbf{x}_{ij} = \mathbf{x}^{(i)} - \mathbf{x}^{(j)}$, and
where $\lb f(x) \rb^{-} = \min\lb 0, f(x) \rb$ ensures that the force acts only when cells overlap (i.e.\ when their
separation is less than $\ell_0$). The cell-cell interaction is symmetric but opposite, meaning $\mathbf{f}_\text{cell-cell}^{(i)} =  - \mathbf{f}_\text{cell-cell}^{(j)}$.
While $\gamma_\text{cc}$ is symmetric across the pair, to preserve generality, we permit asymmetries in cellular properties between the doublets with different values for the frictional drag coefficient and relative force strengths.

\section{Results}

\subsection{Model predicts three-state dynamics for the movement of cell pairs confined to circular micropatterns}
To probe the system response, we place two cells equidistant from the center of the disk-shaped micropattern, induce polarization in an independently chosen random direction, and allow the system to evolve according to Eqs.~\eqref{eq:singlet_forcebalance}, \eqref{eq:polarity_a}, \eqref{eq:polarity_c}, and~\eqref{eq:cell-cellforce} for multiple rotation cycles (Fig.~\ref{fig:doublet_schematic}A). Fig.~\ref{fig:doublet_schematic}B plots one rotational cycle, in either direction, as the pair navigates the confining geometry in our simulation. Along with trajectories, we plot the angular speed of the cell-cell separation plane for sample initializations (Fig.~\ref{fig:doublet_schematic}C). The cell-cell separation plane is defined as the line perpendicular to the axis of radial separation between the cells. The equations of motion are solved numerically using the Forward Euler-Maruyama integration scheme, and parameter values are provided in Table~S1.

The doublets display a variety of motility patterns, including persistent circular motion either clockwise (CW) or counterclockwise (CCW), if the cells act as a cohesive unit, or not moving coherently (NC), if the cells prefer to act as individual units (Fig.~\ref{fig:doublet_schematic}A-B, Movie 1). To classify the emergent behavior, the time evolution of the angular speed is used --- non-coherent movement occurs if the doublet switches directionality over some arbitrary percentage~\footnote{Over 20\% unidirectional for coherent rotation, see SI Table.}, CW for negative angular speeds, or CCW for positive angular speeds. We find the averaged behavior of the system over $3,200$ simulations~\footnote{Without making assumptions about the underlying ratio of a binary proportion ($p=0.5$), and assuming a margin of error $E = \pm3\%$, we require $n = (Z^2\times p \times (1-p)) / E^2 = 1,068$ samples for $95\%$ confidence ($Z=1.96$). Assuming that the lowest ratio of rotation to non-coherent movement never goes below $30\%$, we conclude that $3,200$ samples yield a $95\%$ confidence with error intervals of $\pm 3\%$. Any cases where the ratio of rotation to non-coherent movement falls below $40\%$ will be classified as statistically unreliable.}, and plot the resulting motility state distribution (Fig.~\ref{fig:doublet_schematic}D).


When we combine all 3,200 simulations, we find that the majority of doublets persistently move ($\sim$80\%) in either direction, as indicated by the nearly even distribution between clockwise ($\sim$49\%) and counterclockwise ($\sim$51\%) directions. The marginal case of non-coherent movement appears at a low frequency ($\sim$20\%). At its default parameters, our model predicts three-state behavior --- CW, CCW, or NC --- with the majority of cells persistently rotating either clockwise or counterclockwise with equal transition rates. The three-state dynamics observed in our model are in good agreement with the findings of~\cite{Badih2025} (Fig.~\ref{fig:exp_results}B): when the pair acts as a cohesive unit, the system quickly begins moving in one direction and continues to do so without switching direction or stalling over the course of many hours. However, the lack of a directionality bias is contrary to what was observed in the experimental system in~\cite{Badih2025}, where the CW-bias reached about 60\% of the rotating endothelial doublets.

\subsection{Variability in parameters alone cannot produce bias in rotational movement of the cell doublet}
Can changes to parameter values in the model, including introducing asymmetries in the cellular properties of the doublet system, produce the observed CW bias in~\cite{Badih2025}? 

\begin{figure}[ht!]
    \vspace{-0.25cm}
    \centering
    \includegraphics[width=0.5\textwidth]{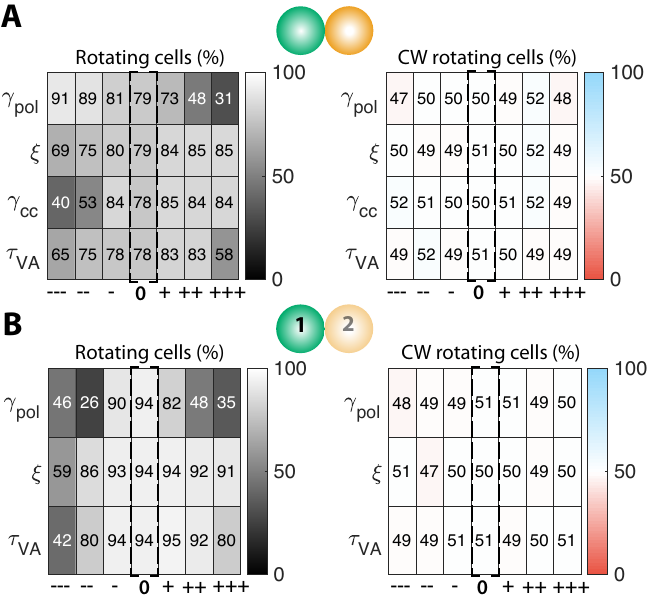}
    \caption{\textbf{Motility outcomes show no directional bias with symmetric or asymmetric parameter regime changes}. Heatmaps showing the percentage of coherent and persistent rotations (left) and CW rotations (right) of cell doublets across model parameter variations (A) in both cells, and (B) in one cell only.}
    \label{fig:doublet_round0_results}
\end{figure}

We probed whether that was the case with changes in either the strength of the polarity force ($\gamma_\text{pol}$), the cell-matrix frictional drag coefficient ($\xi$), the strength of cell-cell adhesion ($\gamma_\text{cc}$), or the response of the velocity alignment mechanism ($\tau_\text{VA}$). Some variations in the parameters did impact the likelihood of rotating coherently -- for example, intuitively, simulated cells were more likely to get stuck or switch direction of movement with decreases in the strength of cell-cell coupling forces, $\gamma_\text{cc}$ (left, Fig.~\ref{fig:doublet_round0_results}A). However, we did not observe a robust directional bias in any of the cases explored (right, Fig.~\ref{fig:doublet_round0_results}A). Asymmetry in the same model parameters of the doublet also had negligible effects (Fig.~\ref{fig:doublet_round0_results}B), regardless of whether the changes were implemented in cell 1 or 2. This indicated that neither mechanical nor biochemical changes, nor heterogeneity alone, can account for the biases in the system. These results held qualitatively with other cell-cell coupling interactions (Fig.~S1-2).

\subsection{Intrinsic bias reveals directionality-tunable rotational movement of doublets}

Next, we wondered what the slightest perturbation to the model could give rise to a bias in directional movement. We extended the polarity model in Eq.~\eqref{eq:polarity_c} by adding a randomly fluctuating intrinsic bias via the term $\mu \mathrm{U}(0,1)$ in the velocity alignment:
\begin{multline}
    \dot{\phi} =
        \frac{1}{\tau_\text{VA}}\bigg[
            \arcsin{\left(\cos{\phi}\sin{\theta_V} - \sin{\phi}\cos{\theta_V}\right)}
                + \mu \mathrm{U}(0,1)\bigg] \\\\
    + \frac{1}{\tau_\text{W}}\arcsin{\left( \cos{\phi}\sin{\theta_W} - \sin{\phi}\cos{\theta_W}\right)}.
\label{eq:polarity_d}
\end{multline}
This is our core assumption about how a preferential cellular organization of the actomyosin cytoskeletal components manifests as an internal torque of the cell body against the underlying matrix. Our motivation is based on the mounting evidence of molecular tilt in the orientation of the cytoskeletal components of some adherent cells~\cite{tee2023actin,yamamoto2025epithelial,mogilner2015cytoskeletal}. 
Here, $\mathrm{U}(0, 1)$ denotes a uniform random variable on $[0, 1]$, so the bias term $\mu \mathrm{U}(0, 1)$
is uniformly distributed over $[\mu, 0]$ where $\mu < 0$ (clockwise) or $[0, \mu]$ when $\mu > 0$ (counterclockwise).
For the default value $\mu = -0.3$ (Table~S1), this corresponds to a random angular perturbation with a mean
of approximately $9$ degrees and a standard deviation of less than $0.1$ degrees.

\begin{figure*}[ht]
    \centering
    \includegraphics[width=\textwidth]{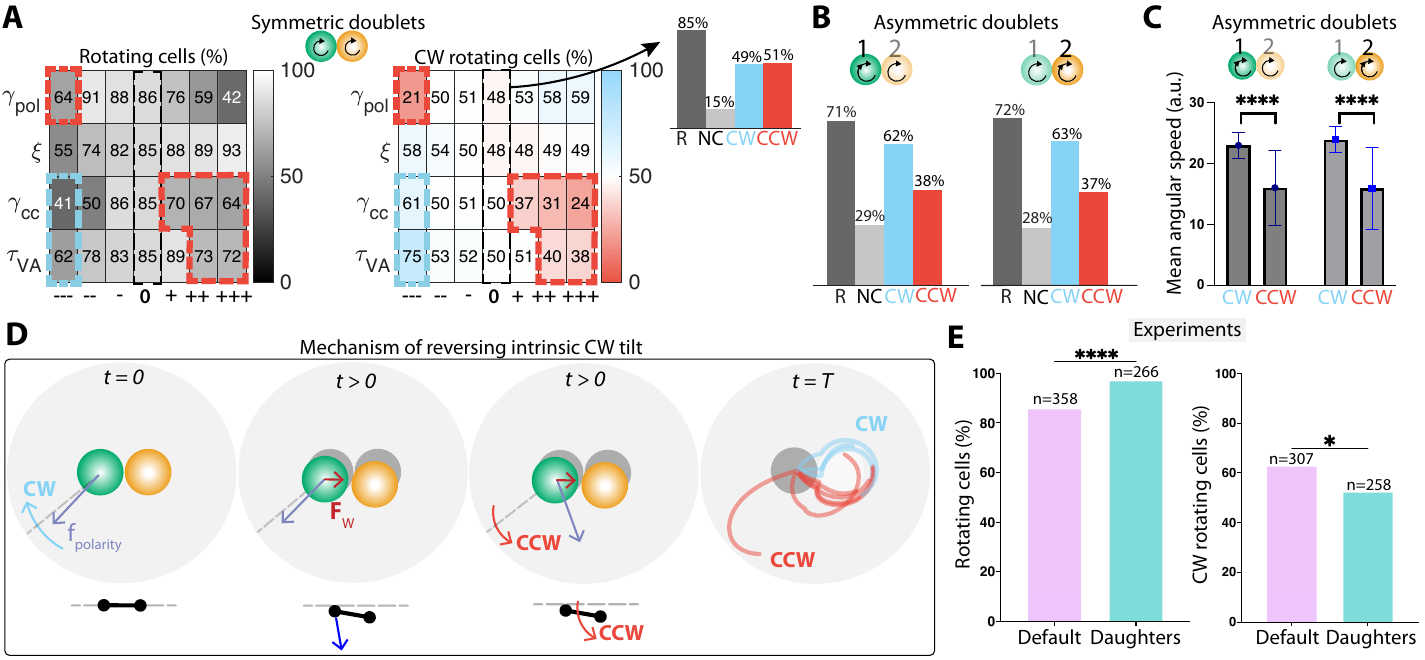}
    \caption{\textbf{By introducing an intrinsic bias, model reproduces persistent circular movement in confinement with a parameter-tunable bias in either direction.} (A) Heatmaps showing the percentage of coherent rotations (left) and CW rotations (right) for parameter sweeps of doublets with intrinsic CW polarity tilt. Dashed black region indicates default values for properties across the pair, blue dashed region marks elevated CW bias ($\geq$60\%), and red dashed region marks elevated CCW bias ($\leq$40\%). Inset: Summary of three-state motility distribution across the dashed black region. (B) Three-state motility distribution and (C) mean angular speed for CW and CCW rotating doublets with asymmetry in velocity alignment timescale in cell 1 (left) or cell 2 (right). The model parameters correspond to an average of the mild parameter difference (-) in Fig.~S4. (D) Schematic of the `tilting dumbbell' mechanism for reversing the CW intrinsic tilt to a CCW bias. (E) Experimental results for the percentage of coherent rotations (left) and CW rotations (right) for HUVEC daughter cell pairs with equal stored mechanical energy. Statistical significance was assessed using an unpaired student's t-test ($^{*}$ indicates $p<0.01$, $^{****}$ indicates $p<0.0001$).}
    \label{fig:doublet_round1_results}
\end{figure*}


The modification revealed a tug-of-war between the effects of centering mechanical forces (confinement and cell-cell coupling) and the biased active polarity force (biochemical polarization) (Fig.~\ref{fig:doublet_round1_results}A). Although the default parameters showed no directionality bias (inset, Fig.~\ref{fig:doublet_round1_results}A), when we repeated the same exploratory parameter search as above, we found that certain parameters produced a CW bias ($\geq 60\%$, blue region in Fig.~\ref{fig:doublet_round1_results}A), while others produced a CCW bias ($\leq 40\%$, red region in Fig.~\ref{fig:doublet_round1_results}A). To obtain a CW bias, the system required a faster (lower) velocity alignment response ($\tau_\text{VA}$) or a weaker cell-cell volume exclusion ($\gamma_\text{cc}$). Our interpretation is that CW bias emerges when cells behave more individualistically, meaning that they have a faster response from their polarization machinery (which is endowed with a mild intrinsic CW bias in the model) compared to the other forces in the system. On the other hand, a CCW bias emerged with weaker polarity forces ($\gamma_\text{pol}$), a slower (higher) velocity alignment response, or a stronger cell-cell volume exclusion. We interpret this as the doublet system moving linked together like a rigid dumbbell, and the applied forces cause a rotation around the center-of-mass, which induces a counteracting torque on the system and yields a CCW bias (Fig.~\ref{fig:doublet_round1_results}D).

While multiple parameter changes could account for the 60\% CW bias observed in the (control) endothelial doublet system, there are two primary reasons we do not consider these perturbations. (1) The cases with CW bias also correspond to instances where the persistence of unidirectional rotational movement is significantly diminished ($\leq62\%$) -- which quantitatively is not in agreement with experimental findings (Fig.~\ref{fig:exp_results}B). (2) These parameter variations are made simultaneously across both cells; however, in~\cite{Badih2025}, it was demonstrated that there is an asymmetry between the cells, specifically in their exerted mechanical energy (Fig.~\ref{fig:exp_results}C, Fig.~4B in~\cite{Badih2025}). Inspired by these findings, we posit that cell-to-cell differences in model parameters could also give rise to a directionality bias.

Indeed, that is the case -- setting an asymmetry in the velocity alignment timescale produced a modest reduction in the average number of rotating doublets from 85\% to 71\%, with 62\% of those persistently rotating doublets in a CW arrangement (Fig.~\ref{fig:doublet_round1_results}B). The results are consistent, independent of which cell has a slightly faster velocity alignment timescale (Fig.~S4). The model also recapitulates the experimental trend in rotational speed --- CW doublets have a significantly higher angular speed compared to CCW doublets (Fig.~\ref{fig:doublet_round1_results}C). Our explanation for the reduction in rotating doublets is that introducing asymmetric polarity response timescales increases the frequency of directional misalignment between cells. These directional conflicts --- where cells' polarities orient in opposite directions --- disrupt the coordination necessary for sustained, cohesive movement. The combination of centering forces from confinement and volume exclusion ensures that the system moves as a dumbbell, with one end that is more responsive to its CW-biased polarity machinery. The trend in speed follows from the same argument: the CW tilt in polarization dynamics preps the system with a directionality that would otherwise have to be agreed upon to resolve head-on collisions (which slow down the coherent motility dynamics).

Qualitatively, the same trends observed with symmetric parameter variations in Fig.~\ref{fig:doublet_round1_results}A were recapitulated with asymmetric changes in cellular properties (Fig.~S4). Namely, a faster timescale in one cell for the re-orientation of the polarity axis in the direction of movement resulted in doublets that preferentially rotated in the CW direction (blue region, Fig.~S4). Taken together, the model informed us that although the observable differences are in the stored mechanical energy of the doublets, the dominant effect is in the timescale of the response of the polarity machinery relative to the remaining force contributions. In other words, in our model, a weaker contractile cell has a more responsive velocity alignment mechanism and is able to quickly re-orient its Rac/Rho polarity axis in the velocity direction. The correlation between the timescale of the velocity alignment response and cellular contractility is not well-established; yet recent work reported that Rho inhibitors increased the timescale of de-adhesion dynamics of human primary keratinocytes~\cite{Srinivasan2019}, interpreted here as a shorter velocity alignment timescale.

\subsection{Experiments with daughter pairs confirm that a lack of asymmetry in cellular properties yields no directional bias}

The model makes two key predictions. (1) There is an intrinsic cellular bias enacted in the polarity machinery, which is assumed to be caused by an internal organization of cytoskeletal components rather than being explicitly captured in this model. (2) By default, cells are primed along the unbiased parameter regime (inset, Fig.~\ref{fig:doublet_round1_results}A), and additional asymmetric or symmetric changes in cellular properties are required to produce a directional bias in either direction. If we start under the assumption of parameters along the unbiased region and if there are no additional differences between the cells in terms of their properties (i.e., contractility), then the cells will rotate persistently in either direction with a higher likelihood of coherent rotations than in the asymmetric case. To test whether this modeling prediction has any grounding in the biological setting, we performed new experiments examining the response of a pair of daughter HUVEC cells in the same 60~$\mu m$ disk-shaped micropattern as in~\cite{Badih2025}. Briefly, we found that our model predicts that asymmetry underpins the emergence of bias, which was indeed recapitulated in the experiments with daughter doublets (Fig.~\ref{fig:doublet_round1_results}E).

We performed experiments on daughter HUVEC cells using a common experimental setup (60~$\mu m$ disks and 17~kPa PAA gels) and assumed that symmetric division on a confined pattern would result in daughter HUVEC cells with equal adhesive properties and thus similar stored mechanical energy~\cite{Gupta2021,Dix2018,Jahan2017,Lesman2014,Thery2006}. To challenge the dynamical impact of asymmetry in the stored mechanical energy, we characterized the rotation in daughter doublets. As before, we identified two classes of collectives: doublets displaying no coherence (NC) while the majority of the doublets do display coherent rotation (left, Fig.~\ref{fig:doublet_round1_results}E). On average, 95\% of rotating cells demonstrated coherent rotation, with no distinguishable bias in directionality. This is congruent with our model predictions for the case of no mechanical asymmetry between the doublets (inset Fig.~\ref{fig:doublet_round1_results}A). For equivalent cells, our model predicts an increased likelihood of coherent rotation (85\%), with roughly 49\% of cells rotating in the CW direction, the same as experiments. This finding not only validates the model and aligns with~\cite{Badih2025} showing that stored energy correlates with motility direction, but also provides confidence to further probe the model's response to other perturbations to cellular properties.

\subsection{Differences in frictional drag and protrusive activity strengthen or reverse the CW rotational bias of doublets}

\begin{figure}[h!]
    \centering
    \includegraphics[width=0.5\textwidth]{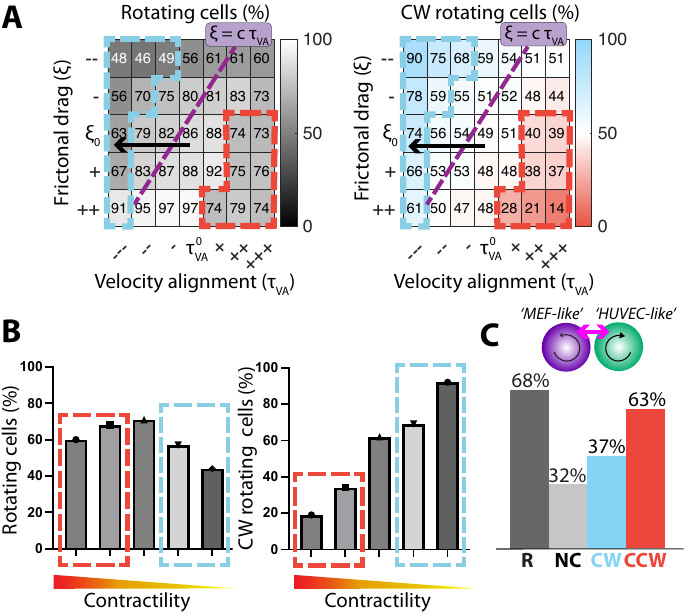}
    \caption{\textbf{Directionality of rotational bias is the result of an interplay between skewed cell polarity and mechanical contributions of frictional drag and cell-cell coupling.} (A) Heatmap of symmetric parameter sweeps in frictional drag coefficient ($\xi$) and polarization timescale ($\tau_\text{VA}$). The dashed purple line marks the 50/50 CW/CCW (unbiased) region. Dashed blue region marks $\geq 60$ CW rotating doublets and dashed red region marks $\geq 40$ CCW doublets. (B) Histogram of the percentage of rotating and CW rotating doublets for parameters choices that correspond to default, high (red) and low (blue) contractility cases. (C) Model results for a simulated heterotypic cell doublet composed of the default CW biased cell (HUVEC-like) together with a CCW biased cell (MEF-like).}
    \label{fig:doublet_biasmodulation}
\end{figure}

Having established that an intrinsic CW bias, together with an asymmetry in the velocity alignment response, can yield the default doublet case (Fig.~\ref{fig:doublet_round1_results}B-C), we next explore which physical forces are modulated by contractility.
We focus on two parameter variations: $\xi$ tunes the strength of the frictional drag against the underlying matrix, and $\tau_{\text{VA}}$ tunes the timescale of alignment of cell polarity with migration direction (Fig.~\ref{fig:doublet_biasmodulation}A).~Mathematically, we uncover an effective bifurcation line $\xi = c \tau_\text{VA}$. A CW bias emerges for $c>c_\text{crit}$ while a CCW bias emerges otherwise, where $c_\text{crit}$ marks no bias, presumably due to a balance of timescale and forces. By tuning model parameters with respect to this bifurcation line, in a way that we will demonstrate is congruent with the experimental perturbations of cellular contractility in Fig.~\ref{fig:exp_results}D-E, we are able to reproduce motility biases similar to those experimentally observed, both in terms of rotational likeliness and directionality (Fig.~\ref{fig:doublet_biasmodulation}B). Parameters used for the contractility modulations are provided in Table S2.

When the timescale of velocity alignment is reduced, resulting in a faster polarity response (top left, Fig.~\ref{fig:doublet_biasmodulation}A), the pair's motility is dominated by their individual intrinsic CW bias, resulting in an overall CW bias. For example, for a fixed parameter choice ($\xi = \xi_0$), decreasing the velocity alignment response timescale from its baseline value ($\tau_\text{VA}^0$) produces a reduction in the number of rotating doublets (from 86\% to 63\%), with a larger percentage of those rotating doublets in a CW arrangement (green arrow, Fig.~\ref{fig:doublet_biasmodulation}A). Similarly, along a fixed drag coefficient $\xi = \xi_0$, increasing the velocity alignment response also reduces the number of rotating doublets, but a CCW bias emerges (73\% of doublets rotating persistently and 39\% of those in the CW direction).
The biases are augmented by additional changes in cell-matrix adhesion; for example, simultaneously decreasing the drag coefficient can result in 90\% of rotating doublets orienting in the CW direction, whereas only 14\% of doublets rotate in the CW direction with slow re-orientation timescales and higher drag coefficient. Lastly, we report that a similar response can be attained by increasing the polarity strength coefficient, $\gamma_\text{pol}$, rather than the velocity alignment timescale. These results also hold when cell-to-cell asymmetry in these parameters is introduced (irrespective of which one of the two cells), rather than homogeneous parameter changes (Fig.~S5B-D).

Next, we connect how these two model parameters --- frictional drag and velocity alignment timescale --- relate to the cytoskeletal contractility machinery that is directly probed in the experiments. Badih et al.\ reported that when doublets were treated with a Rho kinase inhibitor, the system exhibited a more pronounced bias in the CW direction, with an overall lower percentage of persistently rotating doublets (Fig.~\ref{fig:exp_results}D-E). Based on the parameter variations in Fig.~\ref{fig:doublet_biasmodulation}A that produced a stronger CW bias, we posit that the ROCK inhibitor increases the cell's protrusive activity through a faster re-orientation response and a decrease in frictional drag. We found the response to be gradual -- when the velocity alignment timescale is gradually reduced (while maintaining the initial cell-to-cell asymmetry in velocity alignment timescale), CW bias increases while the number of rotating cells decreases (blue region, Fig.~\ref{fig:doublet_biasmodulation}B). In these cases, CW doublets continue to rotate faster than CCW doublets, as was the case for baseline parameters in Fig.~\ref{fig:doublet_round1_results}C. Taken together, these results suggest to us that the addition of the ROCK inhibitor activates the polarization machinery relative to the remaining physical forces in the system such cell-cell interactions or confinement.

What about the control mechanism for flipping the skewness of the rotational orientation towards the CCW direction? The experiments of~\cite{Badih2025} found that Calyculin A, which enhances cellular contractility through increased phosphorylation of myosin light chain, increased the stored mechanical energy of the doublets. More surprisingly, it flipped the prior CW bias in coherent rotations while simultaneously lowering the likelihood of engaging in coherent rotational movement (Fig.~\ref{fig:exp_results}D-E). We interpret increased contractility in the model by making two key assumptions --- a longer timescale for velocity alignment response and a higher frictional drag coefficient (red region, Fig.~\ref{fig:doublet_biasmodulation}B). Our rationale is that contractile cells adhere strongly to the underlying matrix and, thus, have slower dynamics of polarization and motility. Importantly, we find that if we maintain the assumption of cell-to-cell asymmetry in the polarity timescale, we need to increase the volume repulsion strength to decrease the percentage of coherent rotation. This assumption of stronger repulsion is experimentally motivated --- in the Calyculin A experiments, the nuclei appear to be further apart compared to control cases. This leads us to conclude that the Calyculin A treatment diminishes protrusive activity, and thus the contribution of the CW biased intrinsic polarity machinery in exchange for a stronger repulsive interaction between the cells. The slower dynamics constrain the doublet to move in phase-lock, and perturbations off the geometric axis of the doublet yield more frequently to CCW circular motions.

\subsection{The intrinsic cellular bias acts as a destabilizer while the timescale of the polarity re-alignment governs the directionality bias}

In our mathematical model, we find that frictional drag, together with the cell polarity response (strength or velocity alignment timescale), is a proxy for cellular contractility. As a final test of this modeling framework, we ask if we can replicate the heterotypic doublet experiments in~\cite{Badih2025}. A heterotypic doublet system is composed of two different cell types: a strongly contractile mouse embryonic fibroblast (MEF) cell together with an HUVEC. These doublets exhibit not only a high difference in stored mechanical energy but also a clear CCW rotational bias. Important for our model assumptions, the MEF cells have a very different population bias in rotational directionality -- at the population level, MEF cell clusters in a confined disk move persistently in the CCW orientation, opposite to HUVEC clusters, which move in the CW direction.

To test our model, we couple the default HUVEC-like cell to a cell with an opposite intrinsic bias ($\mu>0$, CCW) in the same disk-shaped geometry. Not surprisingly, if all model parameters are the same across the doublet, we find no emergent bias; 81\% of doublets rotate coherently, and 50\% of those are in the CCW direction. However, if we further enforce that MEF cells have a slower (longer) velocity alignment response --- the same assumption as our more contractile (Calyculin A) manipulation in the previous section --- we find that the likelihood of coherent rotation decreases (68\%). Of those cases, 63\% are in the CCW direction, matching qualitatively with experimental findings (Fig.~\ref{fig:doublet_biasmodulation}C). We note that to obtain these results, we also need to increase the strength of cell-cell repulsion, and that at least a two-fold higher value for the velocity alignment timescale is required to reproduce the CCW bias. Further increasing the velocity alignment timescale does yield even more CCW bias and fewer coherently rotating cells. We also found that the absence of a directionality bias for the intrinsic tilt of the MEF cell does not qualitatively change our findings. We intuit that this is because the intrinsic tilt serves as a symmetry break rather than a driver of the underlying dynamical system.

\section{Discussion}
\begin{figure}
    \centering
    \includegraphics[width=0.5\textwidth]{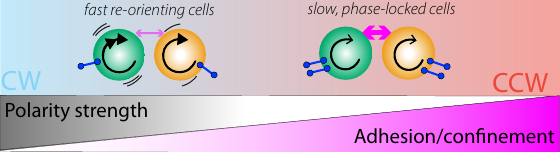}
    \caption{\textbf{Chirality bias is modulated by an interplay between strength of polarity, confinement, and cell-cell interaction forces.} Schematic of working model hypothesis with contractility modulation.}
    \label{fig:doublet_summary}
\end{figure}

Our model constitutes a minimal biophysical framework that identifies the requirements for directional bias in the motility of a two-cell system confined to a disk geometry. To recapitulate the experimentally observed rotational directional bias in HUVEC doublets, two ingredients
are required: (1) individual cells must possess an intrinsic polarity bias, and (2) presence of mechanochemical asymmetry between the cells, specifically asymmetry in the cell polarity timescale (or strength) that governs how quickly cells reorient their polarity axis in the direction of their own velocity. These two assumptions together create a tunable tug-of-war between biased polarity forces and symmetric centering forces from confinement and cell-cell adhesion (Fig.~\ref{fig:doublet_summary}). By modulating the relative strengths and timescales of these competing forces --- which we interpret as changes in cellular contractility --- the system can amplify or reverse its rotational bias, precisely as shown with ROCK inhibition and CalyA treatment. Our framework also generalizes to daughter cell doublets that lack asymmetries in cellular properties, as well as to other cell types.

We first tested whether a movement bias can emerge from such dynamics across the parameter regime, including whether cell-to-cell variability was considered. Parameter variability in the strength of cell polarity, cell-cell interaction, polarity re-orientation timescale, or frictional drag always resulted in a 50/50 CW/CCW directional split (Fig.~\ref{fig:doublet_round0_results}, Fig.~S2). This demonstrates that mechanical or biochemical changes alone cannot break rotational symmetry. A symmetry-breaking mechanism is required. Introducing a mild and random intrinsic CW tilt ($\mu < 0$) in the polarity machinery of each cell breaks symmetry. With symmetric parameter changes across the doublet, some directional preference emerges, but the bias is weak and depends sensitively on parameter choices (Fig.~\ref{fig:doublet_round1_results}A). However, when we introduce doublet asymmetry --- specifically a difference in velocity alignment timescales
($\tau_{VA}^{(1)} \neq \tau_{VA}^{(2)}$) between cells, representing differences in how quickly cells re-orient their polarity axis in the direction of their velocity --- robust and tunable bias emerges. The assumption is also biologically motivated; there is a measurable asymmetry in the stored mechanical energy between doublets reported in~\cite{Badih2025}. Parameter sweeps uncover an effective bifurcation in the $\xi, \tau_{VA}$ plane
(Fig.~\ref{fig:doublet_biasmodulation}): when the polarity response is fast relative to mechanical timescales, CW bias is maintained and even amplified; when it is slow, the system counterintuitively reverses directionality to CCW despite both cells having CW intrinsic bias. The $\tau_{VA}$ heterogeneity creates a ``tilting dumbbell'' where the faster-responding cell ``leads,'' while the slower cell anchors (Fig.~\ref{fig:doublet_round1_results}D), producing the observed $60\%$ CW bias, a reduction in the percentage of coherent rotations, and a faster rotation of CW versus CCW doublets.

Our framework predicts that symmetric doublets should exhibit a high percentage of coherent rotation (95\%) with no directional bias. Experiments with daughter HUVEC pairs --- which lack developed mechanical differences --- confirm this modeling prediction: 95\% of daughter doublets rotate coherently (versus only 5\% exhibiting non-coherent switching), with an equal 50/50 CW/CCW split among the rotating pairs (Fig.~\ref{fig:doublet_round1_results}E). In contrast, non-daughter (asymmetric) pairs show 80\% coherent rotation with a 60\% CW bias among rotating doublets. The finding suggests that tissue-level cell-to-cell variability may be functionally utilized to generate directional information during morphogenesis.

ROCK inhibition (decreased contractility) corresponds in our model to a faster re-orientation timescale and a lower frictional drag coefficient, shifting the system toward high protrusive activity, as others have suggested~\cite{Srinivasan2019,Lomakin2015}. This amplifies the CW bias from
$60\%$ to $74\%$ while reducing persistent rotation from $80\%$ to $63\%$, as faster velocity alignment increases sensitivity to directional perturbations (Fig.~\ref{fig:doublet_biasmodulation}A). Conversely, Calyculin A
(enhanced contractility) corresponds to a slower re-orientation timescale, higher drag, and, critically, stronger cell-cell repulsion --- supported by increased nuclear separation in treated doublets. This multi-parameter shift crosses the bifurcation line, reversing bias to $60\%$ CCW with $55\%$ coherent rotational persistence. The model thus interprets contractility as tuning the balance between Rho-mediated actomyosin contractility and Rac-mediated protrusion, modulating the strength of the polarity machinery relative to mechanical constraints.

Our model makes several simplifying assumptions. Intrinsic bias is represented as a torque rather than arising
from explicit cytoskeletal chirality~\cite{tee2023actin,wangxu2022,yamamoto2025epithelial,Li2022}. Cells are modeled as point particles, neglecting cell shape~\cite{ziebert2012} and 3D actin organization. Velocity alignment is treated phenomenologically rather than from Rho GTPase dynamics~\cite{Mori2008,Buttenschon2020}, and friction is assumed constant rather than mechanosensitive. The relationship between velocity alignment timescale and contractility is postulated rather than derived from first principles. Future models that couple polarization to cytoskeletal architecture, incorporate mechanosensitive feedback, and derive timescales from biochemical networks would test the robustness of our framework.

Other models have shown that torques exchanged through shared cortical interfaces can drive rotation in 3D cell doublets~\cite{Lu2024,Pimpale2020}. In our 2D confined geometry, cell-substrate friction rather than cell-cell cortical coupling is the dominant mechanical interaction, and the brief, repulsive nature of cell-cell contacts suggests that tangential force transmission at the interface plays a lesser role. Incorporating cell-cell torque exchange would be a natural extension for systems with sustained adhesive contacts.

Regarding the choice of intercellular coupling, we demonstrate in the Supporting Material (Fig.~S1-3) that our results hold qualitatively for other coupling forms, specifically passive elastic springs. Lastly, we note that while doublets are observed to be non-rotating experimentally, our simulations instead produce non-coherent rotational movement in which doublets switch their rotational direction. Introducing non-rotation as a true steady state would require additional model ingredients, such as a time-evolving polarity magnitude.

The framework parallels chiral active matter~\cite{tenHagen2014,DiLeonardo2010}, but with dynamic tunability. The principle that small collectives (2-4 cells) occupy a tunable regime while larger groups lock --- may represent an evolved strategy balancing exploration with stable patterning during left-right axis specification. The timescale competition we identify --- between velocity alignment, confinement, and cell-cell interaction --- provides a general mechanism for how cellular collectives integrate multiple mechanical cues, relevant to neural crest migration, epithelial rotation, and cancer invasion. By tuning the intrinsic bias directionality to the population behavior, our model can recapitulate the behavior of other cell types. Importantly, the two requirements identified by our model, intrinsic polarity bias and mechanochemical asymmetry, are not specific to HUVECs. We demonstrate this by recapitulating the CCW bias of heterotypic MEF-HUVEC doublets simply by reversing the sign of the intrinsic bias and adjusting the velocity alignment timescale to reflect higher MEF contractility (Fig.~\ref{fig:doublet_biasmodulation}C). Because the model's ingredients, namely cell polarity, substrate friction, and volume exclusion, are generic features of adherent migrating cells, the framework should be applicable to any system where confined cells exhibit persistent rotational movement. 
Cellular contractility emerges as a master regulator, modulating relative timescales to amplify, erase, or reverse rotational bias.

\textbf{Author Contributions.} A.B. and C.C. designed research; A.B. and C.C. implemented models; E.I., A.B., and C.C. performed simulations; G.B. and L.K. performed experiments; All authors analyzed the data and wrote the paper.


\textbf{Data, Materials, and Software Availability.} Code to reproduce the figures is deposited on Zenodo:~\url{https://zenodo.org/records/17485866}. All other study data are included in the article and/or Supporting Material.

\bibliographystyle{unsrt}
\bibliography{rotationaldoublet.bib}

\end{document}